\documentclass{elsart}

\usepackage{amssymb}

\newcommand{\eq}[1]{Eq.~(\ref{#1})}

\newcommand{\oq}{\frac{1}{4}\,} 
 
\newcommand{\eqs}[1]{Eqs.~(\ref{#1})}

\newcommand{\bref}[1]{Ref.~\cite{#1}}

\renewcommand{\d}{{\rm d}}

\newcommand{\ub}{{\rm ub}}
\newcommand{\bd}{{\rm b}}
\newcommand{\piad}{\pi_{\rm ad}}
\newcommand{\tpi}{\tilde\pi_{\rm ad}}

\newcommand{\va}{v_{\rm b}}
\newcommand{\eps}{\epsilon}
\begin{document}

\begin{frontmatter}

\title{Walks of molecular motors interacting with immobilized filaments}

\author[amst]{Theo M. Nieuwenhuizen},
\author[mpi]{Stefan Klumpp} and
\author[mpi]{Reinhard Lipowsky}

\address[amst]{Instituut voor Theoretische Fysica, Universiteit van Amsterdam, Valckenierstraat~65,
  1018 XE Amsterdam, The Netherlands}
\address[mpi]{Max-Planck-Institut f\"ur Kolloid- und
  Grenzfl\"achenforschung, 14424~Potsdam-Golm, Germany}

\begin{abstract}
Movements of molecular motors on cytoskeletal filaments are described
by directed walks on a line. Detachment from this line is allowed to occur
with a small probability. Motion in the surrounding fluid is described by
symmetric random walks. Effects of detachment and reattachment are
calculated by an analytical solution of the master equation.
Results are obtained for the fraction of bound motors, their average
velocity and displacement. 
Enclosing the system in a finite geometry (tube, slab) leads to an
experimentally realizable problem, that is studied in
 a continuum description and also numerically in a lattice simulation.
\end{abstract}

\begin{keyword}
molecular motors \sep random walks \sep lattice models \sep Brownian motion
\PACS 87.16.Nn \sep 05.40.-a \sep 05.60.-k
\end{keyword}
\end{frontmatter}


\section{Introduction}
\label{sec:intro}

The interior of cells is both highly structured and dynamical. Active
transport therefore plays a crucial role to target molecules to the
various compartments of the cell as well as to maintain and reorganize the
cell's structure. These tasks are accomplished by the cytoskeleton, a
network of protein fibers, and the cytoskeletal motors, proteins which
use the filaments of the cytoskeleton as highways or rails for
directed transport.

Cytoskeletal motors are enzymes which catalyze the hydrolysis of
adenosinetriphosphate (ATP) and, at the same time, use the free energy
released from this reaction to perform mechanical work and convert it
into directed movements. These motor molecules have been studied
extensively during the last decade, both experimentally and
theoretically. The main emphasis has been on the properties of single
motor molecules, their motor mechanisms, and their directed walks
along filaments \cite{Howard2001,Schliwa_Woehlke2003}.

The nanometer size and piconewton forces of the molecular
motors imply that the typical binding energies are of the order of the
thermal energy $k_{\rm B}T$ and that the motor--filament binding can
be overcome by thermal fluctuations. On large time and length scales
($\gg 1$s and $\gg 1\mu$m), motors perform peculiar random walks,
where periods of directed active movements along filaments alternate
with periods of non-directed Brownian motion in the surrounding fluid
after unbinding from a filament
\cite{Ajdari1995,Lipowsky__Nieuwenhuizen2001,Nieuwenhuizen__Lipowsky2002,Nieuwenhuizen__Lipowsky2004}. (A related problem is the effective diffusion of an adsorbed particle along a surface via bulk excursions \cite{Bychuk_OShaughnessy1995}.)
In order to study these random walks, we have recently introduced a
class of lattice models, which are, on the one hand generic in the
sense that they are independent of the specific motor mechanisms, but
which we can, on the other hand, also apply to describe specific motor
molecules by adapting the model parameters to the observed transport
properties \cite{Lipowsky__Nieuwenhuizen2001}. In addition, we can
easily incorporate motor--motor interactions such as the mutual
exclusion from binding sites of the filaments, which leads to 'traffic
jam'-like density pattern and various kinds of phase transitions
\cite{Lipowsky__Nieuwenhuizen2001,Klumpp_Lipowsky2003,Klumpp_Lipowsky2004}

In the following, we will review our results for the motor's random
walks obtained from these lattice models and present some new results
from the corresponding continuum equations. The article is organized
as follows: We introduce the lattice models in section
\ref{sec:model}. In section \ref{sec:2d3d}, we summarize our
analytical results for the random walks on a lattice without confining 
walls. Compartments with confining walls
which are experimentally accessible are studied in section
\ref{sec:finiteGeoms}, where we summarize our numerical results and
present new analytical results from a continuum description of the
random walks.

\section{Lattice models}
\label{sec:model}

We have studied the random walks arising from many encounters with
filaments by mapping them to random walks on a lattice
\cite{Lipowsky__Nieuwenhuizen2001,Nieuwenhuizen__Lipowsky2002}. A line
of lattice sites represents a filament. Motors at these sites perform a biased
random walk and move predominantly in one direction.
Per unit time, they make a forward
step, a backward step and no step with probability $1-\gamma-\delta/2-\epsilon(d-1)/d$, $\delta/2$,
and $\gamma$, respectively, where $d=2,\,3$ is the spatial dimension.
With a small probability $\epsilon/2d$,
they move to each of the adjacent non-filament sites and thus unbind
from the filament. At the non-filament sites the motors perform simple
symmetric random walks and move to each neighbor site with probability
$1/2d$ ($d$ denotes the spatial dimension) and rebind to the filament
with probability $\piad$ when they reach again a filament site.
Confining walls are implemented as repulsive boundaries, at which all
movements into the walls are rejected.

\section{The motors' random walks on a lattice}
\label{sec:2d3d}

Consider a discrete time random walk on a two dimensional square 
lattice with lattice sites labeled by integer coordinates $(n,m)$
with the above transition probabilities.
The master equation for this dynamics reads 
~\cite{Nieuwenhuizen__Lipowsky2002,Nieuwenhuizen__Lipowsky2004}
\begin{eqnarray} \label{me1} 
P_{n,m}(t+1)&=&\oq P_{n+1,m}+\oq P_{n-1,m}+\oq P_{n,m+1}+\oq P_{n,m-1} 
\qquad(m\neq 0,\pm 1)\nonumber\\ \label{me2} 
P_{n,0}(t+1)&=&\oq P_{n,1}+\oq P_{n,-1}+ 
(1-\gamma-\frac{\epsilon+\delta}{2})P_{n-1,0} 
+\frac{\delta}{2} P_{n+1,0}+\gamma P_{n,0}\nonumber\\ 
\label{me4} 
P_{n,\pm 1}(t+1)&=&\oq P_{n+1,\pm 1}+\oq P_{n-1,\pm 1}+\oq P_{n,\pm 2} 
+\frac{\epsilon}{4} P_{n,0}. \label{2dmeq} 
\end{eqnarray}
As initial condition we take an ensemble of particles at $n=m=0$.
Note that, for simplicity, we have chosen the sticking probability $\piad=1$.

The Fourier--Laplace transforms of the probability 
distribution along the filament $P_\bd(n,t)\equiv P_{n,0}(t)$ and of the 
full distribution $P_{n,m}(t)$ are defined as
\begin{equation}\label{p0rz}  
  P_\bd(r,s)\equiv \sum_{t=0}^\infty\sum_{n=-\infty}^\infty 
  \frac{e^{ir n}P_{n,0}(t) }{(1+s)^{t+1}} 
,\quad  
  P(q,r,s)\equiv \sum_{t=0}^\infty\sum_{m,n=-\infty}^\infty 
  \frac{e^{iq m+ir n}P_{n,m}}{(1+s)^{t+1}} 
\end{equation}  
The master equations is reduced to an algebraic equation 
\begin{eqnarray} \label{p2=} 
  P(q,r,s)=\frac{1+[\gamma(1-\cos r) 
    +\frac{1-\eps}{2}(\cos r-\cos q)+i\va \sin r]P_\bd(r,s) } 
  {s+1-\half\cos q-\half\cos r}. 
\end{eqnarray} 
where $\va =1-\gamma-\delta-\half\, \epsilon$ is the average speed on the line.
By integrating this result over $q$ we also obtain $P_\bd(r,s)$ on the 
left hand side. It thus satisfies a linear equation, that is
easily solved. We end up with the distribution
\begin{eqnarray}  
  P_\bd(r,s)&=&
[{s+(1-\gamma)(1-\cos r) 
+\half\eps(\cos r-e^{-\mu})-i\va\sin r}]^{-1} 
 \label{sold=2} 
\end{eqnarray}
for the motors bound to the filament,
where $  \cosh\mu  \equiv  2+2s-\cos r$.
 The distribution 
for the  unbound motors follows from Eq.\ (\ref{p2=}).

\subsection{Properties of the motors bound to the filament line} 
 
{\it Survival fraction. } 
 We extract the transport properties of the motor's 
random walk from (\ref{sold=2}).  The value at $r=0$ 
gives the Laplace transform $N_0(s)$ of the probability 
$N_0(t)\equiv \sum_n P_{n,0}(t)$ that the motor is bound to 
the filament:
\begin{equation} \label{N0s=} 
  N_0(s)=\sum_{t=0}^\infty \frac{N_0(t)}{(1+s)^{t+1}} 
  =
[{(1-\eps)s+\eps\sqrt{s(1+s)}}]^{-1} 
\end{equation} 
For small times this implies
$  N_0(t)\approx 1-2\,{\epsilon \sqrt t}/{\sqrt\pi}+\epsilon^2t,$ 
which is a power series in $\epsilon\sqrt t$.  For $t\ll
1/\epsilon^2$, this is somewhat surprising: although the motors detach
at times $\sim 1/\eps$, the recurrent behavior of the random walk
brings them mostly back to the filament, with half-integer powers in
$t$ due to diffusion.  For times $t\gg 1/\epsilon^2$ we find
$N_0(t)\approx [1-{1}/({2\epsilon^2t})]/ {\sqrt{\pi\,\epsilon^2\,t}}$.
The $ t^{-1/2}$ decay expresses that finally all motors unbind, and
agrees with scaling arguments
~\cite{Ajdari1995,Lipowsky__Nieuwenhuizen2001}.
 
{\it Average position and speed on the filament line. } The expression
(\ref{sold=2}) for $P_\bd(r,s)$ contains much more information.  At
linear order in $r$ we may derive the average position of motor
particles along the filament line, $N_1(t)\equiv\sum_{n}n P_{n,0}(t)$.
We obtain for short times $ N_1(t)= \va t\,[1-{8}{\epsilon \sqrt
  t}/{(3\sqrt\pi)}]$.  The average position of the motors bound to the
filament is given by $ \bar n_\bd(t) \equiv {N_1(t)}/{N_0(t)}\approx
\va t(1-{2}\,{\epsilon \sqrt t}/{3\sqrt\pi})$ and their average speed
is $ \bar v_\bd \equiv {\d \bar n_\bd}/{\d t} \approx\va (1-{ \epsilon
  \sqrt t}/{\sqrt\pi})$, where $\va$ is the average speed if the
particles did not leave the line.  For large times one gets $\bar
n_\bd(t) \approx {\va \sqrt{\pi t}} [1-{2}/({\epsilon\sqrt{\pi
    t}})]/{\epsilon} $ and $ \bar v_\bd(t)\approx{\va \sqrt{\pi}}/{2
  \epsilon\sqrt{t}} =({\pi}/{2})N_0(t)\,\va$, confirming the scaling
$\bar v_\bd(t)\sim \va N_0(t)$
\cite{Ajdari1995,Lipowsky__Nieuwenhuizen2001}.  The effective motor
velocity is reduced by a factor $\sim N_0(t)$, i.e., by the
probability that a motor is in the bound state. The relation $\bar
v_\bd\sim N_0 v_\bd$ also applies to a simple two-state random walk,
where motion is directed in one of the states only. In contrast to the
simple two-state random walk, however, the probability $N_0$ is
time-dependent.
 
{\it Dispersion and diffusion coefficient on the filament line. } From
the second moment $N_2(t)=\sum_n n^2P_{n,0}(t)$ we may define the
normalized second moment $ \overline{n^2}_\bd\equiv{N_2(t)}/{N_0(t)}$,
the dispersion $ \Delta n^2_\bd
\equiv\overline{n^2}_\bd-\overline{n}^2_\bd$ and the diffusion
coefficient, $ D_\bd(t)\equiv(1/2)\,{\d \Delta n^2_\bd}/{\d t}$.  The
results for the latter read at short $t$
\begin{equation}   
  D_\bd(t)
  \approx   \half(1-\gamma)+ 
  \frac{1}{6\sqrt\pi}\,\frac{\va ^2}{\epsilon^2}\,\epsilon^3t^{3/2} 
  +\frac{2\gamma-1}{4}\,\frac{\epsilon\sqrt{t}}{\sqrt{\pi}}. 
\end{equation} 
and for large $t$
\begin{equation}\label{Dparline}  
  D_\bd(t)\approx 
  \frac{\va ^2}{2\epsilon^2} (4-\pi-\frac{\sqrt\pi}{\epsilon\sqrt t}) 
  +\frac{(1-\gamma)\sqrt{\pi}}{4\epsilon\sqrt{ t}}+\frac{1}{4}
\end{equation} 
The limiting value of the diffusion coefficient, $D_\bd(\infty)\sim
(v_\bd/\eps)^2$ is large compared to the diffusion coefficient of the
one-dimensional random walk along the filament,
$D_\bd(0)=(1-\gamma)/2$. This broadening of the distribution occurs
since the unbound motors lag behind the bound ones, which implies that
also the rebinding motors lag behind those that have been bound for
some time.

{\it The density profile on the filament} can be evaluated
analytically for large $n$ and $t$,
\begin{equation} \label{pn0t=} 
  P_{n,0}(t)\approx\frac{\epsilon n}{2\sqrt{\pi \va }(\va t-n)^{3/2}}\, 
  \exp\left(-\frac{\epsilon^2 n^2}{4\va (\va t-n)} \right)
\end{equation} 
The exponential decay for $n\uparrow v_\bd t$ expresses the large
probability of unbinding for $t\gg 1/\eps$.  Comparison with the
results of Monte Carlo simulations shows very good agreement for times
larger than about 8000 time steps \cite{Nieuwenhuizen__Lipowsky2004}.

\subsection{Properties of the unbound motors} 
 
Eventually every motor will unbind and diffuse in the surrounding
fluid. We now discuss the effects of the filament on the behavior of
the unbound motors.

{\it Position and longitudinal diffusion. } One finds for the average
velocity of unbound motors at small times $\bar v_\ub \approx
\frac{2}{3}\va (1-\frac{3}{8}\, \epsilon\sqrt{\pi t})$ and at large
times $ \bar v_\ub(t) \approx {\va }/[\sqrt{\pi}\epsilon\,\sqrt{t}]$.
Whereas each individual motor has zero average velocity in the fluid,
the statistical velocity $\bar v_\ub$ is non-zero, since it is driven
by unbinding from the cloud of motors moving on the filament.  The
longitudinal diffusion coefficient behaves at large times as $
D_\parallel ={(\pi-2)\va ^2}/[{\pi\eps^2}]+{1}/{4}$.  The order of
magnitude $D_\parallel\sim \va ^2/\eps^2$ tells us that, like on the
line, longitudinal diffusion is strongly enhanced by the unbinding
from and binding back to the line.  The perpendicular diffusion
coefficient is modified much less, $
D_\perp(t)\approx{\epsilon\,\sqrt{t}}/({2\sqrt\pi})$ at short times,
while at large times $ D_\perp(t)\approx{1}/{4} -
{1}/{(4\epsilon\sqrt{\pi t}})$.  So the transverse diffusion starts
out very small, and finally reaches its free space value.

{\it The spatio-temporal density profile of the unbound motors} can
also be derived,
\begin{equation} \label{pnmt=} 
  P_{n,m}(t)\approx\frac{\eps(\eps n+2|m|\va )} 
  {2\sqrt{\pi \va }(\va t-n)^{3/2}}\, 
  \exp\left(-\frac{(\eps n+2|m|\va )^2}{4\va (\va t-n)} \right).
\end{equation}

The presence of multiple tracks on microtubules can be modeled by
internal states, which keeps the problem
solvable~\cite{Nieuwenhuizen__Lipowsky2002,Nieuwenhuizen__Lipowsky2004}.

The same setup can be studied in $d=3$
~\cite{Nieuwenhuizen__Lipowsky2002,Nieuwenhuizen__Lipowsky2004}.  The
Fourier-Laplace transformed density on the filament now reads
\begin{equation} P_{\rm b}(r,s)=\frac{3\,I(r,s)}{\epsilon+ 
    [3(1-\epsilon)s+ 
a 
\,(e^{-ir}-1) 
    -
b
(e^{ir}-1)] 
    \,I(r,s)}. 
\end{equation} 
with $a= \half(\epsilon-3\delta)$ and
$b=\half(6-6\gamma-3\delta-5\epsilon)$ and involving the complete
elliptic integral
 $I(r,s)  = (1/2\pi)^2\int_0^{2\pi}{\d q_1}{\d q_2}\, [3+3s-\cos r-\cos q_1-\cos q_2]^{-1}$.
 
The main difference is now that the density of motors on the filament
decays faster, as $\sim1/t$, because of a reduced return probability.
Lack of space prevents us to discuss this any further, but we will
present some results from continuum equations for the
three-dimensional case below.

\section{Motor movements in confined geometries}
\label{sec:finiteGeoms}

\subsection{Scaling arguments and numerical results}

We have also studied these random walks in compartments with simple
geometries which are accessible to {\it in vitro} experiments
\cite{Lipowsky__Nieuwenhuizen2001}. In these geometries, a filament is
immobilized to a surface, and the diffusion of unbound motors is
restricted by confining walls.  In the simplest case, the unbound
motors can diffuse freely in the half space above the surface to which
the filament is immobilized. By placing the filament in a quasi
two-dimensional slab or in a rectangular or cylindrical tube,
diffusion can be restricted along one or two dimensions perpendicular
to the filament. We denote the linear extensions of the compartments
by $L_\perp$.  Scaling arguments and simulations show that, at large
times with $t\gg t_{**}\sim L_\perp^2/D_\ub$, when the
motors experience the presence of the confining walls, these systems exhibit
the same scaling behavior as systems with the same dimensionality, but
without confining walls. For compartments with unconfined diffusion in
$d_\perp$ dimensions perpendicular to the filament, the effective
velocity is given by $v_\bd N_0$ with $N_0(t)\sim t^{-d_\perp/2}$.

\subsection{Continuum equations}

The continuum equations for the random walks of the motors can be
solved in a similar way to the master equations of the lattice models.
Since boundary conditions are easily implemented in the continuum
model, we use them to derive analytical solutions for the half space,
slab and tube geometry, see also \bref{thesis}. We denote
the coordinate parallel to the filament by $x$ 
and the coordinates perpendicular to it by $\mathbf{y}=(y_1,y_2)$.
The continuum equations are given by
\begin{eqnarray}
  \frac{\partial p}{\partial t} & = & D_\ub \Delta p +\delta({\mathbf y})[\tilde\epsilon P-\tpi p_0 ]  \label{diff_gl1_3d}\\
  \frac{\partial P}{\partial t} & = & -v_\bd\frac{\partial P}{\partial x} +D_\bd\frac{\partial^2 P}{\partial x^2}-\tilde\epsilon P +\tpi p_0,\label{diff_gl2_3d}
\end{eqnarray}
with the bound-state velocity and diffusion coefficient $v_\bd$ and
$D_\bd$, the unbound diffusion coefficient $D_\ub$,
$\tilde\epsilon=2\epsilon/3$, $\tpi=2\piad/3$, and $p_0(x)=\ell^2
p(x,\mathbf{y}=\mathbf{0})$.


{\it Full three-dimensional space. } Let us start with the full
three-dimensional space.  We use the Fourier--Laplace transformed form
of \eqs{diff_gl1_3d} and (\ref{diff_gl2_3d}),
\begin{eqnarray}
  s p(r,\mathbf{q},s) & = & -D_\ub (r^2+q_1^2+q_2^2) p(r,\mathbf{q},s) +\tilde\epsilon P(r,s) -\tpi p_0(r,s)\label{diff_gl_transf1_3D}\\
  s P(r,s)  & = & 1+iv_\bd r P(r,s) -D_\bd r^2 P(r,s) -\tilde\epsilon P(r,s)+\tpi p_0(r,s),\label{diff_gl_transf2_3D}
\end{eqnarray}
with the momentum $\mathbf{q}=(q_1,q_2)$ and with the same initial
conditions as before. This leads to
\begin{equation}\label{p0_as_integral_d=3}
  p_0(r,s)  =  \int_{-\infty}^{\infty}\frac{\d q_1}{2\pi}\frac{\d q_2}{2\pi}  p(r,{\mathbf q},s) 
=\frac{\tilde\epsilon P-\tpi p_0}{2\pi D_\ub}\int_{0}^{\infty}\d \tilde q \frac{\tilde q}{1+\tilde q^2}.
\end{equation}
To obtain a convergent integral, we introduce a cutoff $\tilde q_c$ as
the upper integration limit. This cutoff corresponds to a certain
range of attraction or thickness of the filament or a size of the
motor and assures that the motor indeed returns to the filament.
Defining
\begin{equation}
  I\equiv \int_{0}^{\tilde q_c}\d \tilde q \frac{\tilde q}{1+\tilde q^2}=\frac{1}{2}\ln (1+\tilde q_c^2)
 \qquad{\rm with}\qquad \tilde q_c=\sqrt{\frac{D_\ub}{s+D_\ub r^2}}\, q_c,
\end{equation}
we obtain
 $ p_0=\tilde\epsilon P/(2\pi I^{-1}+\tpi)$
 and
\begin{eqnarray}
 P(r,s) & = & \left(s-iv_\bd r +D_\bd r^2+\tilde\epsilon-\frac{\tilde\epsilon \tpi}{2\pi I^{-1}+\tpi}\right)^{-1} \nonumber\\
  & \approx & \left(s-iv_\bd r +\frac{\tilde\epsilon}{\tpi}\frac{4\pi D_\ub}{ \ln(D_\ub q_c^2 s^{-1})}\right)^{-1},
\end{eqnarray}
where the second expression is valid for small $s$ and $r$.  As
before, the binding probability is obtained for $r=0$. For
sufficiently small $s$, the cutoff term can be neglected, and
inverting the Laplace transform leads to
$N_0(t)=P(r=0,t)\approx \tpi/(4\pi \tilde\epsilon D_\ub t)$ in
agreement with the result from the lattice model.


{\it Half space. } The case of the half space can be treated by the
method of reflection and superposition, see e.g.\ \bref{Crank1975}.
The part of the three-dimensional solution for $y_2<0$ can be
reflected at the surface, so that the superposition $p(y_2)+p(-y_2)$
is obtained, which fulfills the boundary condition $\partial
p/\partial{y_2} =0$ at $y_2=0$.  The density profile is symmetric
under reflection at the plane $y_2=0$, so we obtain just twice the
solution for the full three-dimensional space and thus, also the
double for the velocity and the displacement.


{\it Slab. } Let us now consider a slab of height $2L_\perp$, where
the filament is located at $y_1=y_2=0$ in the middle of the slab. The
case where the filament is immobilized at the lower surface of the
slab follows then by reflection and superposition. The boundary
conditions are $\partial p/\partial y_2=0$ at $y_2=\pm L_\perp$ or,
equivalently, periodic boundary conditions at $y_2=\pm L_\perp$. We
can therefore expand the unbound density into a Fourier series,
$p(\dots,y_2)=\sum_{j=-\infty}^{\infty} p(\dots,\omega_j)e^{i\omega_j
  y_2}/(2L_\perp)$ with $\omega_j=\frac{2\pi j}{2L_\perp}$. Here and
in the following we use the symbol $\omega$ for discrete and $q$ for
continuous momentum variables.  For the other variables, we take the
Fourier and Laplace transforms in the same way as above.
\eq{p0_as_integral_d=3} becomes now
\begin{equation}
  p_0(s,r)=\int_{-\infty}^{\infty}\frac{\d q_1}{2\pi} \frac{1}{2L_\perp}\sum_{j=-\infty}^{\infty}p(r,q_1,\omega_j,s).
\end{equation}
We integrate out $q_1$ as before. In addition, we approximate the
summation over $j$ by the term for $j=0$ and twice the integral
\begin{equation}
  S(r,s)\equiv \sum_{j=1}^{\infty} \frac{1}{\sqrt{s+D_\ub r^2+D_\ub \omega_j^2}}  \simeq \frac{L_\perp}{\pi}\int_{\pi/L_\perp}^{\omega_c}\frac{\d \omega }{\sqrt{s+D_\ub r^2+D_\ub \omega^2}},
\end{equation}
where we introduced again a cutoff $\omega_c$. We obtain
\begin{equation}\label{slab_LSG}
  P(r,s)\approx\left[ s-iv_\bd r+\tilde\epsilon-\tilde\epsilon\left(1 + \frac{4L_\perp D_\ub/\tpi}{\sqrt{\frac{D_\ub}{s+D_\ub r^2}}+\frac{2L_\perp}{\pi} \sqrt{D_\ub} S(r,s) }\right)^{-1}\right]^{-1}
\end{equation}
for small $r$.  For motors which have not yet reached the boundaries
of the slab, i.e.\ for $s\gg D_\ub(\pi/L_\perp)^2$, we recover the
result for the full three-dimensional space. For larger times or $s\ll
D_\ub(\pi/L_\perp)^2$, the square root term is dominating in
\eq{slab_LSG} and we obtain
the two-dimensional result \cite{thesis} with an effective
reattachment rate $\tpi/(2L_\perp)$.  In summary, at large times the
probability to be bound to the filament and, thus, the effective
velocity $v_\bd N_0$ behave as
\begin{equation}
  N_0(t)\approx\left\{\begin{array}{cc} \tpi/(4\pi\tilde\epsilon  D_\ub t)     & \quad {\rm for}\quad  t\ll t_{**}\\
      \tpi/(4L_\perp\tilde\epsilon\sqrt{\pi}\sqrt{D_\ub t}) & \quad {\rm for}\quad t\gg t_{**}\end{array} \right.  
\end{equation}
with the crossover time $t_{**}=L_\perp^2/(\pi D_\ub)$.


{\it Tube. } Finally, a {tube with quadratic cross-section} can be
studied in the same way.  In this case we have two discrete momenta
$\omega_{1,i}$ and $\omega_{2,j}$ corresponding to the transverse
coordinates $y_1$ and $y_2$, and \eq{p0_as_integral_d=3} is now
replaced by
\begin{equation}
  p_0(s,r)=\frac{1}{(2L_\perp)^2}\sum_{i,j=-\infty}^{\infty}p(r,\omega_{1,i},\omega_{2,j},s).
\end{equation}
As in the case of the slab, we approximate the double sum by the term
for $i=j=0$ plus an integral with a cutoff $\omega_c$,
\begin{equation}\label{tube_trennung}
  p_0(s,r)=\frac{\tilde\epsilon P-\tpi p_0}{(2L_\perp)^2}\left(\frac{1}{D_\ub r^2+s} +\frac{L_\perp^2}{D_\ub \pi^2}\int_{\pi/L_\perp}^{\omega_c}\frac{2\pi \omega\,\, \d\omega}{D_\ub(r^2+\omega^2)+s}\right),
\end{equation}
which leads to
\begin{equation}
  P(r,s)\approx\left\{ s-iv_\bd r+\frac{4L_\perp^2\tilde\epsilon}{\tpi}\left(\frac{1}{s}+\frac{L_\perp^2}{\pi D_\ub}\ln\frac{1+D_\ub \omega_c^2/s}{1+D_\ub \pi^2/L_\perp^2 s}\right)^{-1}\right\}^{-1}
\end{equation}
for small $s$ and $r$. For $s\gg D_\ub(\pi/L_\perp)^2$, i.e., for
motors which have not yet reached the boundaries, this expression
yields again the solution for the three-dimensional case without
confining walls, and for larger times, it leads to
\begin{equation}
  P(r,s)\approx\left[s(1+4L_\perp^2\frac{\tilde\epsilon}{\tpi})-iv_\bd r\right]^{-1}
\end{equation}
and, therefore, to a constant probability to be bound to the filament,
$N_0(t)\approx 1/(1+4L_\perp^2\tilde\epsilon/\tpi)$, and a constant
effective velocity
$v=v_\bd /(1+4L_\perp^2\tilde\epsilon/\tpi)$.  The crossover to the
long-time regime is governed by the crossover time
$t_{**}=v_\bd/v\times \tpi/(\tilde\epsilon 4\pi D_\ub)\approx
L_\perp^2/(\pi D_\ub)$ where the last expression holds for large
$L_\perp$ and agrees with the crossover time obtained for the slab.

\section{Summary}

Molecular motors exhibit peculiar random walks which arise from the
repeated binding to and unbinding from filaments. We have obtained
analytical results for these random walks in systems without and with
confining walls by solving the master equations of lattice models and
the corresponding continuum equations. Such random walks are relevant
both for intracellular phenomena such as, for example, the 'slow
transport' in axons \cite{Wang__Brown2000} and for artificial systems
in nanotechnology in which molecular motors are used as transporters
\cite{Hess_Bachand_Vogel2004}.

\end{document}